\documentclass[preprint,3p]{elsarticle}
\usepackage{amssymb}
\usepackage{amsthm}
\usepackage[mathlines]{lineno}
\usepackage{graphicx}
\usepackage{units}
\usepackage{url}
\usepackage{amsmath}
\usepackage{amsfonts}
\usepackage{bm}
\usepackage{textcomp}
\usepackage{subfigure}
\usepackage{multicol}
\usepackage{verbatim}
\usepackage{rotating}
\usepackage[colorlinks,linkcolor=red,citecolor=red]{hyperref}
\usepackage{breakurl}
\usepackage{float}
\usepackage{epsfig}
\usepackage{dcolumn}
\usepackage{bm}
\usepackage{color}
\usepackage{pstricks}
\usepackage{pst-node}
\usepackage{times}
\usepackage{indentfirst}
\usepackage[english]{babel}
\biboptions{numbers,sort&compress}
\addto{\captionsenglish}{%

}
\journal{Physics Letters B}

\usepackage{setspace}%

\begin{document}
\begin{frontmatter}
\title{{\bf
\boldmath Observation of the decay $\Lambda_c^+\rightarrow \Sigma^- \pi^+\pi^+\pi^0$}}

\author{
\begin{small}
\begin{center}
M.~Ablikim$^{1}$, M.~N.~Achasov$^{9,e}$, S.~Ahmed$^{14}$, M.~Albrecht$^{4}$, A.~Amoroso$^{50A,50C}$, F.~F.~An$^{1}$, Q.~An$^{47,a}$, J.~Z.~Bai$^{1}$, O.~Bakina$^{24}$, R.~Baldini Ferroli$^{20A}$
% ...
      \\
      \vspace{0.2cm}
      (BESIII Collaboration)\\
      \vspace{0.2cm} {\it
$^{1}$ Institute of High Energy Physics, Beijing 100049, People's Republic of China\\
% ...
$^{4}$ Bochum Ruhr-University, D-44780 Bochum, Germany\\
% ...
$^{9}$ G.I. Budker Institute of Nuclear Physics SB RAS (BINP), Novosibirsk 630090, Russia\\
% ...
$^{14}$ Helmholtz Institute Mainz, Johann-Joachim-Becher-Weg 45, D-55099 Mainz, Germany\\
% ...
$^{20}$ (A)INFN Laboratori Nazionali di Frascati, I-00044, Frascati, Italy; (B)INFN and University of Perugia, I-06100, Perugia, Italy\\
% ...
$^{24}$ Joint Institute for Nuclear Research, 141980 Dubna, Moscow region, Russia\\
% ...
$^{47}$ University of Science and Technology of China, Hefei 230026, People's Republic of China\\
% ...
$^{50}$ (A)University of Turin, I-10125, Turin, Italy; (B)University of Eastern Piedmont, I-15121, Alessandria, Italy; (C)INFN, I-10125, Turin, Italy\\
% ...
$^{a}$ Also at State Key Laboratory of Particle Detection and Electronics, Beijing 100049, Hefei 230026, People's Republic of China\\
% ...
$^{e}$ Also at the Novosibirsk State University, Novosibirsk, 630090, Russia\\
% ...
}\end{center}
\vspace{0.4cm}
\end{small}
}


\begin{abstract}
We report the first observation of the decay $\Lambda^+_{c}\rightarrow \Sigma^- \pi^+\pi^+\pi^0$,
based on data obtained in $e^+e^-$ annihilations with an integrated luminosity of 567~pb$^{-1}$ at
$\sqrt{s}=4.6$~GeV.
The data were collected with the BESIII detector at the BEPCII storage rings.
The absolute branching fraction  $\mathcal{B}(\Lambda^+_{c}\rightarrow\Sigma^-\pi^+\pi^+\pi^0)$
is determined to be $(2.11\pm0.33({\rm stat.})\pm0.14({\rm syst.}))\%$.
In addition, an improved measurement of $\mathcal{B}(\Lambda^+_{c}\rightarrow\Sigma^-\pi^+\pi^+)$
is determined as $(1.81\pm0.17({\rm stat.})\pm0.09({\rm syst.}))\%$.
\\
\\
\text{Keywords:~~branching fraction, charmed baryon, weak decays, $e^+e^-$ annihilation, BESIII }
\end{abstract}

\end{frontmatter}

\begin{multicols}{2}

\section{Introduction}
The study of hadronic decays of charmed baryons provides important information to understand
both the strong and the weak interactions~\cite{1601.04241}. It also provides
essential input to understand background contributions in the study of $b$-baryon physics, as $\Lambda_b$ decays dominantly to $\Lambda_c^+$.
More than 30 years have passed since the $\Lambda^+_c$ baryon was first observed in $e^+e^-$ annihilations by
the Mark~II experiment~\cite{prl44_10} and the knowledge of
$\Lambda^+_c$ decays remains very poor compared to that for charmed mesons. So far, measured decay modes account for only about 60\%~\cite{pdg2014} of all $\Lambda_c^+$ decays, primarily consisting of modes with a $\Lambda(\Sigma)$ hyperon or a proton in the final state. Decays to the $\Sigma^-$ hyperon are Cabibbo-allowed and are expected to have large rates. However, no experimental measurements exist except for $\Lambda_c^+\rightarrow \Sigma^{-}\pi^+\pi^+$~\cite{pdg2014}.
Therefore, searching for additional decay modes with $\Sigma^{-}$ in the final state is important to build up knowledge on $\Lambda_c^+$ decays.
In this paper, we report the first observation of the so-far undetermined, but expected to be large, decay of $\Lambda_c^+\rightarrow \Sigma^-\pi^+\pi^+\pi^0$~\cite{charge}. In addition, we perform the first absolute measurement of the branching fraction for $\Lambda_c^+\rightarrow \Sigma^-\pi^+\pi^+$.

The data analyzed in this work corresponds to an integrated luminosity of $567$~pb$^{-1}$~\cite{lum} of $e^+e^-$ annihilations at center-of-mass energy (c.m.) $\sqrt{s}=4.6$~GeV by the BEPCII collider and collected with the BESIII detector~\cite{Ablikim:2009aa}. The c.m.~energy is slightly above the threshold for the production of
$\Lambda_c^{+}\bar{\Lambda}_c^{-}$, so $\Lambda_c^{+}\bar{\Lambda}_c^{-}$ pairs are produced with no additional hadrons.
The analysis technique in this work, which was first applied in the
Mark~III experiment~\cite{prl62_1821}, is optimized for measuring charm hadron pairs produced near threshold.
First, we select the subset of our events in which a $\bar{\Lambda}_c^-$ is reconstructed in an exclusive
hadronic decay mode, designated as the single-tag (ST) sample. Events in this ST sample are then
searched for the signal channel $\Lambda_c^+\rightarrow \Sigma^-\pi^+\pi^+(\pi^0)$ in the system
recoiling against the ST to select double tag (DT) events.
In the final states of $\Lambda_c^+\rightarrow \Sigma^-\pi^+\pi^+(\pi^0)$, the $\Sigma^-$ hyperon is detected through $\Sigma^-\rightarrow n\pi^-$. As the neutron is not reconstructed in this analysis, we deduce its kinematic properties by four-momentum conservation.
The absolute branching fraction (BF) of $\Lambda_c^+\rightarrow \Sigma^-\pi^+\pi^+(\pi^0)$ is derived from the
probability of detecting the DT signals in the ST sample.
Hence, this method provides a clean and straightforward BF
measurement that is independent of the number of
$\Lambda^+_c\bar{\Lambda}^-_c$ events produced.

\section{BESIII Detector and Monte Carlo Simulation}
BESIII~\cite{Ablikim:2009aa} is a cylindrical detector with
a coverage of 93\% of the full $4\pi$ solid angle. It consists of a Helium-gas based main drift chamber (MDC), a plastic
scintillator time-of-flight (TOF) system, a CsI~(Tl) electromagnetic
calorimeter (EMC), a superconducting solenoid providing a 1.0\,T
magnetic field, and a muon detection system in the iron flux return of the magnet. The charged particle momentum
resolution is 0.5\% at a transverse momentum of 1~GeV/$c$. The
photon energy resolution at 1\,GeV is 2.5\% in the central barrel region and 5.0\% in the two end caps.
More details about the design and performance of the detector are given in
Ref.~\cite{Ablikim:2009aa}.

A GEANT4-based~\cite{geant4} Monte Carlo (MC) simulation package,
which includes the geometric description of the detector and the
detector response, is used to determine the detection efficiency and
to estimate the potential backgrounds. MC samples of the signal mode $\Lambda_c^+\rightarrow \Sigma^-\pi^+\pi^+(\pi^0)$, together with a
$\bar{\Lambda}^-_c$ decaying to specified ST modes, are generated
with KKMC~\cite{kkmc} and EVTGEN~\cite{nima462_152}, taking into account initial-state radiation (ISR)~\cite{SJNP41_466} and final-state radiation~\cite{plb303_163} effects. The $\Lambda_c^+\rightarrow \Sigma^-\pi^+\pi^+(\pi^0)$ decay is simulated by reweighting the phase-space-generated MC events to approximate observed kinematic distributions in data.
To understand potential background contributions, an inclusive MC sample is used. It includes generic $\Lambda_c^+\bar{\Lambda}_c^-$
events, $D_{(s)}^{(*)}\bar{D}_{(s)}^{(*)}+X$ production, ISR return to the charmonium states at lower masses and continuum $q\bar{q}$ processes.
Previously measured decay modes of the $\Lambda_c$, $\psi$ and $D_{(s)}$ are simulated with EVTGEN, using BFs
from the Particle Data Group (PDG)~\cite{pdg2014}. The unknown decays of the $\psi$
states are generated with LUNDCHARM~\cite{lundcharm}.

\section{Analysis}
The ST and DT selection technique that is used in our analysis
follows closely the one used and described in Ref.~\cite{bes3lamev}.
We reconstruct the $\bar{\Lambda}^-_c$ baryons in the eleven hadronic
decay modes listed in Table~\ref{tab:deltaE_1}.
Intermediate particles are reconstructed through their decays
$K^0_S\rightarrow \pi^+\pi^-$, $\bar{\Lambda}\rightarrow
\bar{p}\pi^+$, $\bar{\Sigma}^0\rightarrow \gamma\bar{\Lambda}$ with
$\bar{\Lambda}\rightarrow \bar{p}\pi^+$, $\bar{\Sigma}^-\rightarrow
\bar{p}\pi^0$, and $\pi^0\rightarrow \gamma\gamma$.
The selection criteria for the proton, kaon, pion, $\pi^0$, $K^0_S$ and $\bar{\Lambda}$ candidates
used in the reconstruction of the ST signals are described in Ref.~\cite{bes3lamev}.

\begin{table*}[htp]
\centering
\caption{\label{tab:deltaE_1}\normalsize Requirements on $\Delta E$ and ST yields
$N_{\bar{\Lambda}_c^-}$ for the eleven ST
modes. The uncertainties are statistical only.}
\begin{tabular}{llr} \hline \hline Mode~&~~~$\Delta E$~(GeV)~~~&$N_{\bar{\Lambda}_c^-}$ \\
\hline
 $\bar{p} K^0_S$                & $[-0.025, 0.028]$ &   $1066\pm33$  \\
 $\bar{p} K^+\pi^-$             & $[-0.019, 0.023]$ &   $5692\pm88$  \\
 $\bar{p}K^0_S\pi^0$            & $[-0.035, 0.049]$ &  ~~$593\pm41$  \\
 $\bar{p} K^+\pi^-\pi^0$        & $[-0.044, 0.052]$ &   $1547\pm61$  \\
 $\bar{p} K^0_S\pi^+\pi^-$      & $[-0.029, 0.032]$ &  ~~$516\pm34$  \\
 $\bar{\Lambda}\pi^-$           & $[-0.033, 0.035]$ &  ~~$593\pm25$  \\
 $\bar{\Lambda}\pi^-\pi^0$      & $[-0.037, 0.052]$ &   $1864\pm56$  \\
 $\bar{\Lambda}\pi^-\pi^+\pi^-$ & $[-0.028, 0.030]$ &  ~~$674\pm36$  \\
 $\bar{\Sigma}^0\pi^-$          & $[-0.029, 0.032]$ &  ~~$532\pm30$  \\
 $\bar{\Sigma}^-\pi^0$          & $[-0.038, 0.062]$ &  ~~$329\pm28$  \\
 $\bar{\Sigma}^-\pi^+\pi^-$     & $[-0.049, 0.054]$ &   $1009\pm57$  \\
\hline \hline
\end{tabular}
\end{table*}

The ST $\bar{\Lambda}^-_c$ signals are identified using the beam-energy-constrained mass, $M_{\rm BC}=\sqrt{E^2_{\rm
beam}-|\vec{p}_{\bar{\Lambda}^-_c}|^2}$, where $E_{\rm
beam}$ is the beam energy and
$\vec{p}_{\bar{\Lambda}^-_c}$ is the momentum of the
$\bar{\Lambda}^-_c$ candidate in the rest frame of the initial $e^+e^-$ system~\cite{frame:comment}. To improve the signal purity, the
energy difference $\Delta E=E_{\rm beam}-E_{\bar{\Lambda}^-_c}$ for
each candidate is required to be within approximately
$\pm3\sigma$ of the $\Delta E$ signal peak position, where
$\sigma$ is the $\Delta E$ resolution and
$E_{\bar{\Lambda}^-_c}$ is the reconstructed $\bar{\Lambda}^-_c$
energy. Table~\ref{tab:deltaE_1} shows the mode-dependent
$\Delta E$ requirements and the ST yields in the $M_{\rm BC}$ signal
region $(2.280, 2.296)$~GeV/$c^2$,
which are obtained by fits to the $M_{\rm BC}$ distributions.
See Ref.~\cite{bes3lamev} for more details.
The total ST yield is
$N^{\rm tot}_{\bar{\Lambda}^-_c}=14415\pm159$, where the uncertainty is statistical only.

Candidates for the decay $\Lambda^+_c\rightarrow \Sigma^-\pi^+\pi^+(\pi^0)$ with $\Sigma^-\rightarrow n\pi^-$ are
reconstructed from the tracks not used in the ST
$\bar{\Lambda}^-_c$ reconstruction.
It is required that there are only three charged
tracks in the system recoiling against the $\bar{\Lambda}_c^-$ satisfying
$|\cos\theta|<0.93$, where $\theta$ is the polar angle with respect to the beam direction. For the two $\pi^+$ candidates from the $\Lambda_c^+$, the distances of
closest approach to the interaction point must be
within $\pm10$\,cm along the beam direction and within 1\,cm in the
perpendicular plane, while the $\pi^-$ candidate from $\Sigma^-$ decay is not subjected to this requirement. Identification of charged tracks is
performed by combining the $dE/dx$ information from the MDC and the time of flight measured in the TOF to obtain the probability $\mathcal{L}_{h}$ for each hadron type $h$. The three charged pions must satisfy $\mathcal{L}_{\pi} > \mathcal{L}_{K}$.
Photon candidates are reconstructed from isolated clusters in the
EMC in the regions $|\cos\theta| \le 0.80$ (barrel) and $0.86 \le
|\cos\theta|\le 0.92$ (end cap). The deposited energy of a neutral cluster is required to be larger than 25 (50)~MeV in the
barrel (end cap) region, and the angle between the photon candidate
and the nearest charged track must be larger than 10$^\circ$. To
suppress electronic noise and energy deposits unrelated to the
event, the difference between the EMC time and the event start time
is required to be within $(0, 700)$~ns.
To reconstruct $\pi^0$ candidates, the invariant mass of photon pairs is required to be within (0.110, 0.155)~GeV/$c^2$ and, as a second step, a
kinematic fit is implemented to constrain the $\gamma\gamma$ invariant mass to the nominal $\pi^0$ mass~\cite{pdg2014}.

The kinematic variable
\footnotesize
$$M_n=\sqrt{(E_{\rm
beam}-E_{\pi^+\pi^+\pi^-(\pi^0)})^2-|\overrightarrow{p}_{\Lambda_c^+}-\overrightarrow{p}_{\pi^+\pi^+\pi^-(\pi^0)}|^2}$$
\normalsize
is computed to characterize the
reconstructed mass of the undetected neutron, where $E_{\pi^+\pi^+\pi^-(\pi^0)}$ is the energy of the $\pi^+\pi^+\pi^-(\pi^0)$ combination and  $\overrightarrow{p}_{\pi^+\pi^+\pi^-(\pi^0)}$ is the three-momentum of the $\pi^+\pi^+\pi^-(\pi^0)$ combination.
The expected momentum $\vec{p}_{\Lambda_c^+}$ of the $\Lambda_c^+$ is calculated by
$\vec{p}_{\Lambda_c^+}=-\hat{p}_{\rm tag}\sqrt{E_{\rm beam}^2-m^2_{\Lambda^+_c}},$
where $\hat{p}_{\rm tag}$ is the direction of the momentum of the
ST $\Bar{\Lambda}_c^-$ candidate and $m_{\Lambda_c^+}$ is the mass
of the $\Lambda^+_c$ taken from the PDG~\cite{pdg2014}.
Similarly, we can construct the variable
\footnotesize
$$M_{n\pi^-}=\sqrt{(E_{\rm
beam}-E_{\pi^+\pi^+(\pi^0)})^2-|\overrightarrow{p}_{\Lambda_c^+}-\overrightarrow{p}_{\pi^+\pi^+(\pi^0)}|^2}$$
\normalsize
to represent the reconstructed mass of the $\Sigma^-$.

\begin{figure*}[htp]
\centering
   \begin{minipage}[t]{8.0cm}
   \includegraphics[height=7cm,width=8.6cm]{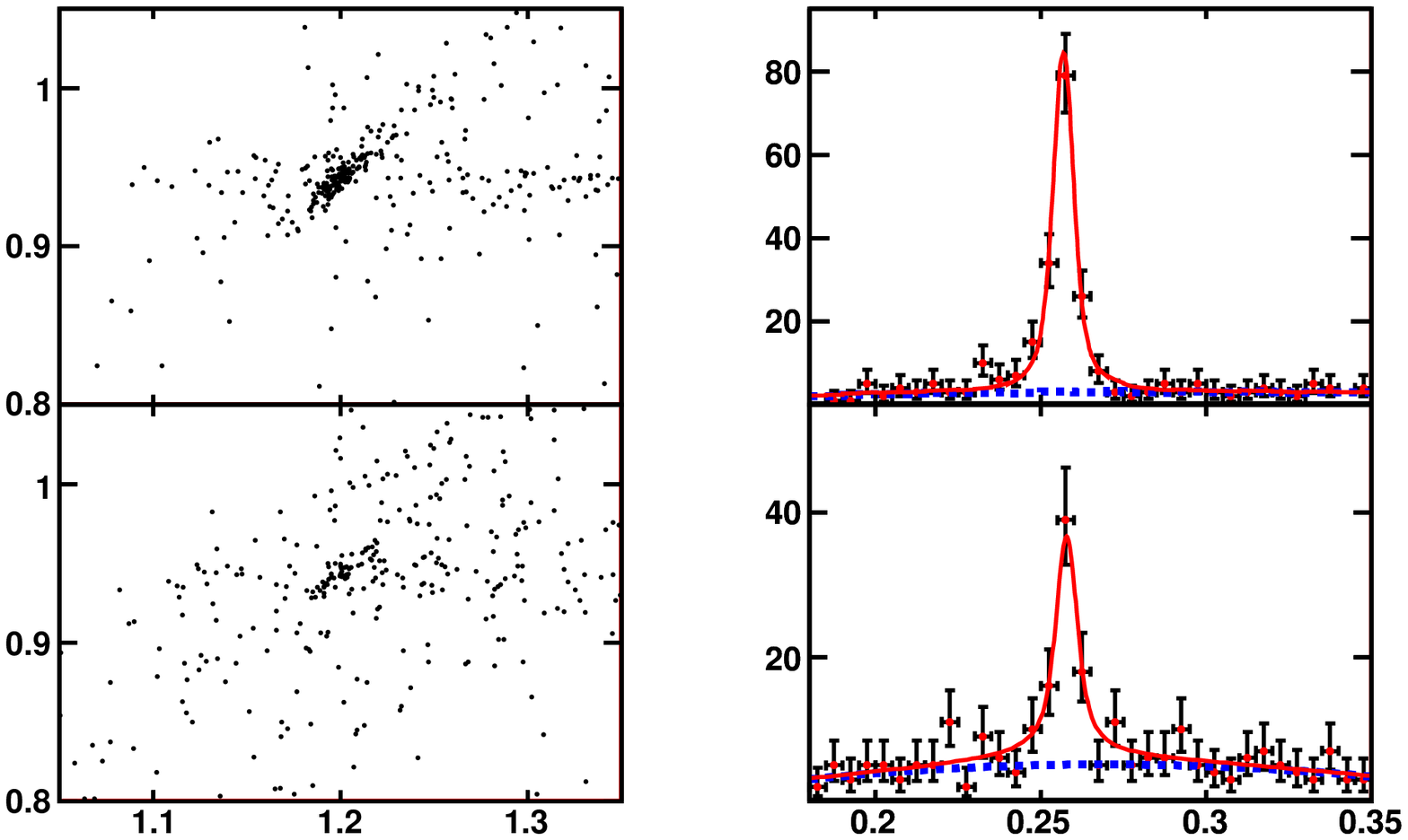}
   \put(-155,175){\bf  (a)}
   \put(-210,3){\bf \small $M_{n\pi^-}$ (GeV/$c^2$) }
   \put(-245,80){\rotatebox{90}{\bf \small $M_{n}$ (GeV/$c^2$)}}
   \put(-35, 175){\bf  (c)}
   \put(-100,3){\bf \small $M_{n\pi^-}-M_n$ (GeV/$c^2$) }
   \put(-120,70){\rotatebox{90}{\bf \small Events/0.005 GeV/$c^2$}}
   \put(-155,90){\bf   (b)}
   \put(-35, 90){\bf   (d)}
   \end{minipage}
   \caption{ \normalsize  Scatter plots of $M_n$ versus $M_{n\pi^-}$ for candidates in data for (a) $\Lambda_c^+\rightarrow \Sigma^-\pi^+\pi^+$ and (b) $\Lambda_c^+\rightarrow \Sigma^-\pi^+\pi^+\pi^0$. Also shown are fits to the distributions of $M_{n\pi^-}-M_n$ for (c) $\Lambda_c^+\rightarrow\Sigma^-\pi^+\pi^+$ and (d) $\Lambda_c^+\rightarrow\Sigma^-\pi^+\pi^+\pi^0$ in data. Solid lines are the results of a complete fit
while dashed lines reflect the background components.} \label{fig:scat_data}
\end{figure*}

The distributions of $M_{n}$ versus $M_{n\pi^{-}}$ for the $\Lambda^+_c\to
\Sigma^-\pi^+\pi^+$ and $\Lambda^+_c\to
\Sigma^-\pi^+\pi^+\pi^0$ candidates in data are
shown in Figs.~\ref{fig:scat_data} (a) and (b), respectively, where clusters corresponding to signal decays are evident.
To improve the resolution of the signal mass, as well as to better handle the backgrounds around the $\Sigma^-$ and neutron mass regions, we determine the signal yields from the distribution of the mass difference $M_{n\pi^-}-M_n$, since $M_{n\pi^-}$ and $M_n$ are highly correlated. Based on a study of the inclusive MC samples, no peaking backgrounds are expected for these two channels.
We perform an unbinned maximum likelihood fit to the $M_{n\pi^-}-M_n$ spectra, as shown in Figs.~\ref{fig:scat_data} (c) and (d).
In the fits, the signals are described by non-parametric functions
extracted from the signal MC convoluted with a Gaussian function accounting for the resolution difference between data and MC, while the background shapes are described with a second-order polynomial function. The width of the Gaussian is left free in the fit, while its mean is fixed to zero. From the fits, we find the DT signal yields $N^{\rm obs}_{\Sigma^-\pi^+\pi^+}=161\pm15$ and
$N^{\rm obs}_{\Sigma^-\pi^+\pi^+\pi^0}=88\pm14$, where the uncertainties are statistical only.
Backgrounds from non-$\Lambda_c^+$ decays are estimated by examining
the ST candidates in the $M_{\rm BC}$ sideband $(2.252, 2.272)$~GeV/$c^2$ in data. The backgrounds from
non-$\Lambda_c^+$ decays are found to be negligible.

The absolute BFs for $\Lambda_c^+\rightarrow \Sigma^-\pi^+\pi^+$ and $\Lambda_c^+\rightarrow \Sigma^-\pi^+\pi^+\pi^0$ are determined by
\begin{eqnarray}
&&\mathcal{B}(\Lambda_c^+\rightarrow \Sigma^-\pi^+\pi^+(\pi^0)) \nonumber \\
&&=\frac{N^{\rm obs}_{\Sigma^-\pi^+\pi^+(\pi^0)}}{N^{\rm tot}_{\bar{\Lambda}_c^-}\cdot\varepsilon_{\Sigma^-\pi^+\pi^+(\pi^0)}\cdot\mathcal{B}(\Sigma^-\rightarrow n\pi^-)},~~~~~~~
\label{eq:branch}
\end{eqnarray}
where $\varepsilon_{\Sigma^-\pi^+\pi^+(\pi^0)}$ is the detection efficiency for
the $\Lambda_c^+\rightarrow \Sigma^-\pi^+\pi^+(\pi^0)$ decay with $\Sigma^-\to n \pi^-$.
The intermediate decay branching fraction of $\Sigma^-\to n \pi^-$ is included in the denominator of Eq.~\eqref{eq:branch}.
For each ST mode $i$, the efficiency $\varepsilon^i_{\Sigma^-\pi^+\pi^+(\pi^0)}$ is
obtained by dividing the DT efficiency $\varepsilon^i _{{\rm tag}, \Sigma^-\pi^+\pi^+(\pi^0)}$ by
the ST efficiency $\varepsilon^i _{\rm tag}$.
After weighting $\varepsilon^i_{\Sigma^-\pi^+\pi^+(\pi^0)}$ by the mode-by-mode ST yields in data, we find the overall average efficiencies $\varepsilon_{\Sigma^-\pi^+\pi^+}=(61.8\pm0.4)\%$ and $\varepsilon_{\Sigma^-\pi^+\pi^+\pi^0}=(29.0\pm0.2)\%$, where the branching fraction for $\pi^0\rightarrow \gamma\gamma$ is included.
Substituting the values of $N^{\rm obs}_{\Sigma^-\pi^+\pi^+(\pi^0)}$, $N^{\rm tot}_{\bar{\Lambda}^-_c}$, $\varepsilon_{\Sigma^-\pi^+\pi^+(\pi^0)}$ and $\mathcal{B}(\Sigma^-\rightarrow n\pi^-)$ in
Eq.~(\ref{eq:branch}),
we obtain $\mathcal{B}(\Lambda^+_{c}\rightarrow\Sigma^-\pi^+\pi^+)=(1.81\pm0.17\pm0.09)\%$
and $\mathcal{B}(\Lambda^+_{c}\rightarrow\Sigma^-\pi^+\pi^+\pi^0)=(2.11\pm0.33\pm0.14)\%$, where the first uncertainties are
statistical, and the second are systematic, as described below.

With the DT technique, the BF measurement is insensitive to uncertainty in the ST efficiencies.
The systematic uncertainties in measuring
$\mathcal{B}(\Lambda_c^+\rightarrow \Sigma^-\pi^+\pi^+)$ and $\mathcal{B}(\Lambda_c^+\rightarrow \Sigma^-\pi^+\pi^+\pi^0)$ mainly arise from
the efficiencies of $\pi$ detection and identification, fits to the $M_{n\pi^-}-M_n$ distributions and the signal modelling in the MC simulation.
The systematic uncertainties in the $\pi^{\pm}$ tracking and identification are both
determined to be 1.0\% by studying a set of samples of
$e^+e^-\rightarrow \pi^+\pi^-\pi^+\pi^-$, $e^+e^-\rightarrow
K^+K^-\pi^+\pi^-$ and $e^+e^-\rightarrow p\bar{p}\pi^+\pi^-$ obtained from data with c.m.\ energy above 4.0~GeV. The $\pi^0$ reconstruction efficiency
is validated by analyzing DT events with $\bar{D}^0\rightarrow K^+\pi^-$ or $K^+\pi^-\pi^0$ versus $D^0\rightarrow K^-\pi^+\pi^0$~\cite{pi0eff}. The difference of the $\pi^0$ reconstruction efficiencies between data and MC simulations is estimated to be 2.0\%.
The uncertainty from the fit to the $M_{n\pi^-}-M_n$ distribution is evaluated by checking the relative changes of $N^{\rm obs}_{\Sigma^-\pi^+\pi^+(\pi^0)}$ with different choices for signal shapes (double Gaussian function), background shapes (first-order polynomial function, third-order polynomial function and a MC-derived background shape) and fit ranges ((0.19, 0.34)~GeV/$c^2$).
The uncertainty in modelling the signal process is obtained by varying the reweighting factors of the observed kinematic variables within their
statistical uncertainties and extracting the difference of the resultant efficiencies. The difference is estimated to be 2.0\% for the studied channels and
is taken as the systematic uncertainty due to the signal modelling.
In addition, there are systematic uncertainties in obtaining $N^{\rm tot}_{\bar{\Lambda}_c^-}$ evaluated by using alternative signal shapes in the fits to the
$M_{\rm BC}$ spectra~\cite{bes3lamev}, resulting in an uncertainty of 1.0\%, and in the statistical limitation of the MC samples, which is estimated to be 0.6 (0.7)\% for $\Lambda_c^+\rightarrow \Sigma^-\pi^+\pi^+(\pi^0)$. The uncertainties from the BFs of $\Sigma^-\rightarrow n\pi^-$ and $\pi^0\rightarrow \gamma\gamma$ are negligible. All of the above systematic uncertainties
are summarized in Table~\ref{tab:syst}, and the total
uncertainties are evaluated to be 5.2\% and 6.4\% for $\mathcal{B}(\Lambda_c^+\rightarrow \Sigma^-\pi^+\pi^+)$ and $\mathcal{B}(\Lambda_c^+\rightarrow \Sigma^-\pi^+\pi^+\pi^0)$, respectively, by combining all items in quadrature.

\begin{table*}
\centering
\caption{\normalsize  \label{tab:syst}Summary of the relative systematic uncertainties $\Delta^{\rm syst}_{\Sigma^-\pi^+\pi^+}$ and $\Delta^{\rm syst}_{\Sigma^-\pi^+\pi^+\pi^0}$ in $\mathcal{B}(\Lambda^+_c\rightarrow \Sigma^-\pi^+\pi^+)$ and $\mathcal{B}(\Lambda^+_c\rightarrow \Sigma^-\pi^+\pi^+\pi^0)$, respectively. } \normalsize
\begin{tabular}
{l|c|c} \hline Source & $\Delta^{\rm syst}_{\Sigma^-\pi^+\pi^+}$ [\%] & $\Delta^{\rm syst}_{\Sigma^-\pi^+\pi^+\pi^0}$ [\%] \\
\hline
$\pi^{\pm}$ tracking                  & 3.0  & 3.0  \\
$\pi^{\pm}$ identification            & 3.0  & 3.0  \\
$\pi^0$ reconstruction                & $\cdots$ & 2.0  \\
Fit to $M_n-M_{n\pi^-}$               & 2.0  & 3.6  \\
Signal modelling                      & 2.0  & 2.0  \\
MC statistics                         & 0.6  & 0.7  \\
$N^{\rm tot}_{\bar{\Lambda}_c^-}$     & 1.0  & 1.0  \\
\hline
Total                         & 5.2  & 6.4  \\
\hline
\end{tabular}
\end{table*}

\section{Summary}

Based on an $e^+e^-$ collision data sample with an integrated luminosity of 567~pb$^{-1}$ taken at
$\sqrt{s}=4.6$~GeV with the BESIII detector, we report the first
observation of the decay $\Lambda^+_c\rightarrow \Sigma^-\pi^+\pi^+\pi^0$ and the first absolute BF measurement for
$\Lambda^+_{c}\rightarrow
\Sigma^-\pi^+\pi^+$. The results are $\mathcal{B}(\Lambda^+_{c}\rightarrow\Sigma^-\pi^+\pi^+)=(1.81\pm0.17\pm0.09)\%$
and $\mathcal{B}(\Lambda^+_{c}\rightarrow\Sigma^-\pi^+\pi^+\pi^0)=(2.11\pm0.33\pm0.14)\%$, where the first uncertainties are statistical and the second are systematic.

Our result for $\mathcal{B}(\Lambda^+_{c}\rightarrow\Sigma^-\pi^+\pi^+)$ is consistent with and more precise than the previous result~\cite{pdg2014}.
BESIII measured the BF of the isospin symmetric channel $\mathcal{B}(\Lambda^+_c\rightarrow \Sigma^+\pi^+\pi^-)=(4.25\pm0.24\pm0.20)\%$~\cite{prl116_052001}. This allows us to determine the ratio $\mathcal{B}(\Lambda^+_c\rightarrow \Sigma^-\pi^+\pi^+)/\mathcal{B}(\Lambda^+_c\rightarrow \Sigma^+\pi^+\pi^-)=0.42\pm0.05\pm0.02$, where the first uncertainty is statistical and the second systematic. The statistical uncertainty of the ratio dominates, as many common systematic uncertainties cancel.
This is consistent with and more precise than the value previously measured by the E687 Collaboration $(0.53\pm0.15\pm0.07)$~\cite{Frabetti:1994kt}.

\section{Acknowledgments}
The BESIII collaboration thanks the staff of BEPCII and the IHEP computing center for their strong support. This work is supported in part by National Key Basic Research Program of China under Contract No. 2015CB856700; National Natural Science Foundation of China (NSFC) under Contracts Nos. 11125525, 11235011, 11275266, 11305180,
11322544, 11322544, 11335008, 11425524, 11505010; the Chinese Academy of Sciences (CAS) Large-Scale Scientific Facility Program; the CAS Center for Excellence in Particle Physics (CCEPP); Joint Large-Scale Scientific Facility Funds of the NSFC and CAS under Contracts Nos.~U1332201, U1532257, U1532258; CAS under Contracts Nos. KJCX2-YW-N29, KJCX2-YW-N45, QYZDJ-SSW-SLH003; 100 Talents Program of CAS; National 1000 Talents Program of China; INPAC and Shanghai Key Laboratory for Particle Physics and Cosmology; German Research Foundation DFG under Contracts Nos. Collaborative Research Center CRC 1044, FOR 2359; Istituto Nazionale di Fisica Nucleare, Italy; Joint Large-Scale Scientific Facility Funds of the NSFC and CAS; Koninklijke Nederlandse Akademie van Wetenschappen (KNAW) under Contract No. 530-4CDP03; Ministry of Development of Turkey under Contract No.~DPT2006K-120470; National Science and Technology fund; The Swedish Resarch Council; U.S.\ Department of Energy under Contracts Nos. DE-FG02-05ER41374, DE-SC-0010118, DE-SC-0010504, DE-SC-0012069; University of Groningen (RuG) and the Helmholtzzentrum fuer Schwerionenforschung GmbH (GSI), Darmstadt; WCU Program of National Research Foundation of Korea under Contract No. R32-2008-000-10155-0. This paper is also supported by Beijing municipal government under Contract Nos. KM201610017009, 2015000020124G064.


%
\end{multicols}


\begin{thebibliography}{99}
\bibitem{charge} Throughout this paper, charged conjugate modes are implied unless explicitly stated otherwise.
\bibitem{frame:comment} All kinematic quantities presented in this paper are evaluated in the rest frame of the initial $e^+e^-$ system.
\end{thebibliography}
\end{document}